\documentclass{jps-cp}
\usepackage{txfonts} 
\usepackage{amsmath}
\usepackage{amssymb}
\usepackage{bm}

\title{RKKY Interaction and Quadrupole Order in Pr$T_2$Al$_{20}$ ($T$=Ti, V) Based on Effective 196 Orbital Model Extracted from First-Principles Calculation}

\author{Yuto \textsc{Iizuka}$^{1}$, Takemi \textsc{Yamada}$^{2}$, Katsurou \textsc{Hanzawa}$^{2}$, Yoshiaki \textsc{\=Ono}$^{1}$}

\inst{$^{1}$Department of Physics, Niigata University, Niigata, 950-2181, Japan

$^{2}$Department of Physics, Faculty of Science and Technology, Tokyo University of Science, Chiba, 278-8510, Japan
}

\email{y.iizuka.phys@gmail.com}

\recdate{September 17, 2019}

\abst{
Electronic states and quadrupole orders in the 1-2-20 systems Pr$T_{2}$Al$_{20}$ ($T$=Ti, V) are investigated on the basis of the first-principles band calculation. As the de Haas-van Alphen experiments reveal that the Pr-4$f$ electrons in the systems are sufficiently localized and irrelevant for the Fermi surface, we derive the low-energy effective tight-binding models consists of 196 orbitals of conduction electrons so as to reproduce the first-principles electronic structures of La$T_{2}$Al$_{20}$ ($T$=Ti, V) without contribution from the 4$f$ electrons. 
Based on the effective models, we calculate the Ruderman-Kittel-Kasuya-Yosida (RKKY) interaction between the quadrupole moments of the Pr ions mediated by the conduction electrons. The obtained results indicate that the wave vector of the expected quadrupole order is ${\bm{Q}}=\left(0,0,0\right)$ for PrTi$_{2}$Al$_{20}$ while it is ${\bm{Q}}=\left(\pi/a,\pi/a,0\right)$ for PrV$_{2}$Al$_{20}$ as consistent with experimental observations in PrTi$_{2}$Al$_{20}$ and PrV$_{2}$Al$_{20}$ which exhibit ferro- and antiferro-quadrupole orders, respectively. 
}

\kword{Pr 1-2-20 systems, RKKY Interaction, quadrupole order, first-principles calculation}

\begin{document}
\maketitle

\section{Introduction}

Recently, the Pr 1-2-20 systems Pr$T_2 X_{20}$ have been intensively investigated concerning their specific features including quadrupole orders, superconductivity and non-Fermi liquid behaviors [1]. The crystalline electric field ground states of the Pr ions are the non-Kramers doublets $\Gamma_{3}(\Gamma_{23})$ which have the electric quadrupole moments with no magnetic moment. In fact, PrTi$_2$Al$_{20}$ and PrV$_2$Al$_{20}$ exhibit the ferro-quadrupole (FQ) and the antiferro-quadrupole (AFQ) orders, respectively [2,3]. They also show remarkable superconductivities coexisting with the quadrupole orders [4]. To discuss the quadrupole order and the superconductivity, the details of the energy bands and the Fermi surfaces are important. In the Pr 1-2-20 systems, the de Haas-van Alphen experiments revealed that the Fermi surfaces are well accounted for by the first-principles band calculations for the corresponding La 1-2-20 systems [5,6], indicating that the Pr 4$f$ electrons are sufficiently localized. Therefore, the quadrupole orders are considered due to the Ruderman-Kittel-Kasuya-Yosida (RKKY) interaction between the quadrupole moments of the localized Pr 4$f$ electrons. The purpose of this study is to evaluate the RKKY interaction on the basis of the realistic energy band structure extracted from the first-principles calculation [7] and to discuss the quadrupole orders in PrTi$_2$Al$_{20}$ and PrV$_2$Al$_{20}$.

\section{Effective 196 Orbital Model}

\begin{figure}[tbh]
\begin{center}
\includegraphics[width=80mm]{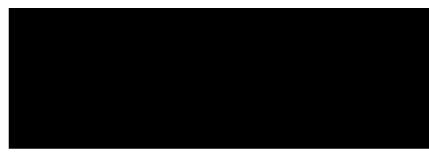}\vspace{3mm}
\includegraphics[width=80mm]{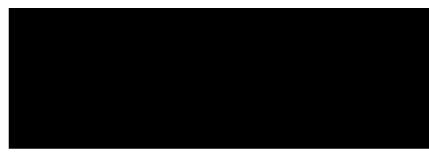}\vspace{1mm}
\caption{The energy band dispersions from the 196 orbital model for LaTi$_{2}$Al$_{20}$ (a) and LaV$_{2}$Al$_{20}$ (b) together with those from WIEN2k, where the Fermi level is set to be zero.}
\label{f1}
\end{center}
\end{figure}

First, we perform the first-principles band calculation for La$T_{2}$Al$_{20}$ ($T$=Ti, V) by using WIEN2k code on the basis of the density-functional theory (DFT) with the generalized gradient approximation (GGA), where $17\times17\times17$ ${\bm k}$-points, the muffin-tin radius $R_{\rm MT} = 2.5$ (2.3) a.u. for La and $T$ (Al) and the the plane-wave cuttoff $K_{\rm max}=3.2$ (a.u.)$^{-1}$ are used. We also use the experimentally determined lattice parameters of Pr$T_{2}$Al$_{20}$ ($T$=Ti, V) [8] instead of La$T_{2}$Al$_{20}$ in order to discuss the former electronic states in the 4$f$ electron localized regime. In addition, we employ the GGA+U method with $U=60$eV to exclude the La-4$f$ components near the Fermi level. The obtained Fermi surfaces of LaTi$_{2}$Al$_{20}$ are in good agreement with those in the previous study [6] which well accounts for the dHvA experiment in PrTi$_{2}$Al$_{20}$ as mentioned before. Then, we construct the tight binding models for the conduction electrons so as to reproduce the electronic structures near the Fermi level for La$T_{2}$Al$_{20}$ ($T$=Ti, V) as shown in Figs. 1 (a) and (b), respectively,  by using the maximally localized Wannier functions which consist of 196 orbitals: La-$d$ (5 orbitals $\times$ 2 sites), La-$s$ (1 orbital $\times$ 2 sites), $T$-$d$  (5 orbitals $\times$ 4 sites), $T$-$s$  (1 orbital $\times$ 4 sites), Al-$p$ (3 orbitals $\times$ 40 sites) and Al-$s$ (1 orbital $\times$ 40 sites) in the face-centered cubic (fcc) unit cell.

\section{RKKY Interaction and Quadrupole Order}

The RKKY interaction between the localized $f$ electrons is the indirect interaction mediated by the $c$ electrons via the hybridization between the $c$ and $f$ electrons. 
As the Pr $5d$ electrons give the dominant contribution to the susceptibility responsible for the RKKY interaction between the quadrupoles of the Pr ions as will be shown later, we consider the $c$-$f$ hybridization only between the $5d$ and $4f$ on the same Pr atom which becomes finite for the CEF Hamiltonian ${\cal H}_{\rm CEF}$ without inversion symmetry [9] and is explicitly given by
\begin{align}
V_{mm'l \sigma }^{\left( n \right)} &= \left\langle {{f^n}m } \right|{{\cal H}_{\rm CEF}}\left| {{f^{n - 1}}m'} \right\rangle \left| {l \sigma } \right\rangle,
\end{align}
where $\left| {{f^{n}}m} \right\rangle$ is a state of the Pr $4f^n$ configuration labeled by $m$ and $\left| {l \sigma } \right\rangle$ is a Pr $5d$ state with the orbital $l$ and spin $\sigma$, where the $t_{2g}$ orbital exclusively hybridizes with the $f$ orbital in the case with the $T_{\rm d}$-symmetric CEF Hamiltonian. The second-order perturbation with respect to $V_{mm'l \sigma }^{\left( n \right)}$ yields the Kondo exchange interaction whose Hamiltonian is given by
\begin{align}
{\cal H}_{\rm ex} =  \sum\limits_{\gamma \gamma'}^{} \sum\limits_{l,l'}^{} \sum\limits_{\sigma \sigma '}^{} K^{\gamma \gamma'}_{ll'\sigma \sigma '}
\left| l\sigma \right\rangle \left| f^2 \Gamma _{3}(\gamma') \right\rangle \left\langle f^2 \Gamma _{3} ( \gamma) \right| \left\langle l'\sigma ' \right|
\end{align}
with the Kondo coupling
\begin{align}
\label{Jk}
K^{\gamma \gamma'}_{ll'\sigma \sigma '} 
& = 
 \sum\limits_m^{} {\left( {\frac{{V_{\gamma ml'\sigma '}^{\left( 2 \right)}V_{l\sigma m\gamma '}^{\left( 2 \right)}}}{{\varepsilon _{{t_{2g}}}^d + E_m^{\left( 1 \right)} - E_{{\Gamma _{3}}}^{\left( 2 \right)}}} + \frac{{V_{l\sigma m\gamma '}^{\left( 2 \right)}V_{\gamma ml'\sigma '}^{\left( 2 \right)}}}{{\varepsilon _{{t_{2g}}}^d + E_m^{\left( 1 \right)} - E_{{\Gamma _{3}}}^{\left( 2 \right)}}}} \right)} \nonumber \\
&+
 \sum\limits_m^{} {\left( {\frac{{V_{m\gamma 'l'\sigma '}^{\left( 3 \right)}V_{l\sigma \gamma m}^{\left( 3 \right)}}}{{E_m^{\left( 3 \right)} - E_{{\Gamma _{3}}}^{\left( 2 \right)} - \varepsilon _{{t_{2g}}}^d}} + \frac{{V_{l\sigma \gamma m}^{\left( 3 \right)}V_{m\gamma 'l'\sigma '}^{\left( 3 \right)}}}{{E_m^{\left( 3 \right)} - E_{{\Gamma _{3}}}^{\left( 2 \right)} - \varepsilon _{{t_{2g}}}^d}}} \right)}, 
\end{align}
where $E_m^{\left(n \right)}$, $E_{{\Gamma _{3}}}^{\left( 2 \right)}$ and $\varepsilon _{t_{2g}}^d$ are the energies for the state $m$ of the Pr $4f^n$ with $n=1,3$, the $\Gamma _{3}$ ground state of the Pr $4f^2$ and the $t_{2g}$ state of the Pr $5d$, respectively. From the second-order perturbation with respect to $K^{\gamma \gamma'}_{ll'\sigma \sigma '}$, we obtain the RKKY interaction Hamiltonian for the $\Gamma_{3}$-type electric quadrupoles $O_{u}\propto \frac{1}{2}\left(3J_{z}^{2}-\bm{J}^{2}\right)$ and $O_{v}\propto \frac{\sqrt{3}}{2}\left(J_{x}^{2}-J_{y}^{2}\right)$ of the Pr ions as
\begin{align}
\label{Hrkky_multipoles}
{{\cal H}}_{{\rm{RKKY}}} =  - \sum\limits_{\langle ij \rangle}^{}\left( 
J_{ij}^{O_u} O_u^i O_u^j + J_{ij}^{O_v} O_v^i O_v^j
\right)
\end{align}
with the RKKY interactions for $O_u$ and $O_v$ between the Pr ions at the positions ${\bm{r}}_{i}$ and ${\bm{r}}_{j}$
\begin{align}
\label{Jrkky_OuOu}
&J_{ij}^{O_u} = \frac{1}{{64}}\left( J_{ij}^{1111} - J_{ij}^{1122} - J_{ij}^{2211} + J_{ij}^{2222} \right), \\
\label{Jrkky_OvOv}
&J_{ij}^{O_v} = \frac{1}{{64}}\left( J_{ij}^{1212} + J_{ij}^{1221} + J_{ij}^{2112} + J_{ij}^{2121} \right), 
\end{align}
where the RKKY interaction in  the orbital representation is given by
\begin{align}
\label{Jrkky}
J_{ij}^{{\gamma _1}{\gamma _2}{\gamma _3}{\gamma _4}} = \frac{1}{{N}}\sum\limits_{\bm{q}} {\sum\limits_{\left\{ l  \right\}} {\sum\limits_{\left\{ \sigma  \right\}} {K^{{\gamma _1}{\gamma _2}}_{{l _1}{l _2}{\sigma _1}{\sigma _2}}K^{{\gamma _3}{\gamma _4}}_{{l _4}{l _3}{\sigma _2}{\sigma _1}}} } }
\chi _{{l _1}{l _2}{l _3}{l _4}}^{}\left( {\bm{q}} \right)e^{i{\bm{q}} \cdot \left({\bm{r}}_{i}-{\bm{r}}_{j}\right)}
\end{align}
with the wave vector $\bm{q}$, the total number of the unit cell $N$ and the bare susceptibility for the $c$ electrons in the orbital representation given by 
\begin{align}
\chi _{{l _1}{l _2}{l _3}{l _4}}^{}\left( {\bm{q}} \right)
= \frac{1}{N}\sum\limits_{s s'}^{} \sum\limits_{\bm{k}}^{} 
 \frac{{f({\varepsilon _{{\bm{k}} + {\bm{q}}s'}^{}} ) - f( {\varepsilon _{{\bm{k}}s }^{}} )}}{{\varepsilon _{{\bm{k}}s }^{} - \varepsilon _{{\bm{k}} + {\bm{q}}s'}^{}}}
u_{{l _1}s }^ * \left( {\bm{k}} \right)u_{{l _2}s'}^{}\left( {{\bm{k}} + {\bm{q}}} \right) u_{{l _3}s }^{}\left( {\bm{k}} \right)u_{{l _4}s'}^ * \left( {{\bm{k}} + {\bm{q}}} \right) ,
\end{align}
where $\varepsilon _{\bm{k}s}$ is the energy for the $c$ electrons with the wave vector $\bm{k}$ and the band $s$ and $u_{l s}\left(\bm{k}\right)$ is the corresponding eigen vector with the orbital $l$, and $f(\varepsilon)$ is the Fermi distribution function.

Substituting  $\varepsilon _{\bm{k}s}$ and  $u_{l s}\left(\bm{k}\right)$ obtained from the 196  orbital model into eq. (8), we calculate the RKKY interactions eqs. (5) and (6) with eq. (7), where we set $T=0.01$eV and the unknown parameters included in the Kondo coupling eq. (3) such as the Coulomb interaction between the 4$f$ electrons and the strength of the CEF potential are assumed so as to reproduce the transition temperature of the FQ (AFQ) for PrTi$_2$Al$_{20}$ (PrV$_2$Al$_{20}$) within the mean-field approximation for the RKKY Hamiltonian eq. (2). 
Figs. 2 (a) and (b) show the RKKY interactions $J_{ij}$ for PrTi$_2$Al$_{20}$ and PrV$_2$Al$_{20}$, respectively. For both compounds, $J_{ij}$ exhibit oscillatory decreases with increasing the relative distance over ten unit cells. The nearest neighbor coupling is positive (ferroic) for PrTi$_2$Al$_{20}$ (see Fig. 2 (a)), while it is negative (antiferroic) for PrV$_2$Al$_{20}$ (see Fig. 2 (b)). 

To explicitly determine the wave vector $\bm{q}$ of the expected quadrupole order, we perform the Fourier transformation of the RKKY interaction $J_{ij}$. Fig. 3 (a) and (b) show the $\bm{q}$-dependence of the Fourier transformed RKKY interactions $J(\bm{q})$ for LaTi$_{2}$Al$_{20}$ and LaV$_{2}$Al$_{20}$, respectively. 
We find that  $J(\bm{q})$ has a maximum at ${\bm{q}}=\left(0,0,0\right)$ for PrTi$_{2}$Al$_{20}$ (see Fig. 3 (a)),  while it does at ${\bm{q}}=\left(\pi/a,\pi/a,0\right)$ for PrV$_{2}$Al$_{20}$ (see Fig. 3 (a)) as consistent with experimental observations in PrTi$_{2}$Al$_{20}$ and PrV$_{2}$Al$_{20}$ which exhibit FQ and AFQ orders, respectively. The difference in the $\bm{q}$-dependence of $J(\bm{q})$ between the two compounds is mainly due to the difference in the band dispersions near the Fermi level and the Fermi surfaces: there are small Fermi surfaces around the $\Gamma$ point for PrTi$_{2}$Al$_{20}$ (see Fig. 1 (a)) while not for PrV$_{2}$Al$_{20}$ (see Fig. 1 (b)). We also perform the first-principles band calculation for La$T_{2}$Zn$_{20}$ ($T$=Ir, Rh) and find that there is no small Fermi surface around the $\Gamma$ point similar to the case with PrV$_{2}$Al$_{20}$. This seems to be consistent with the experiments where Pr$T_{2}$Zn$_{20}$ ($T$=Ir, Rh) show the AFQ orders [1]. To be more conclusive, we need explicit calculations based on the effective multi-orbital models for those compounds. It is also interesting to discuss the possible superconductivity mediated by the orbital fluctuations which can be discussed on the basis of the RKKY Hamiltonian. Such calculations are under the way and will be reported in subsequent papers.

\begin{figure}[tbh]
\begin{center}
\includegraphics[width=90mm]{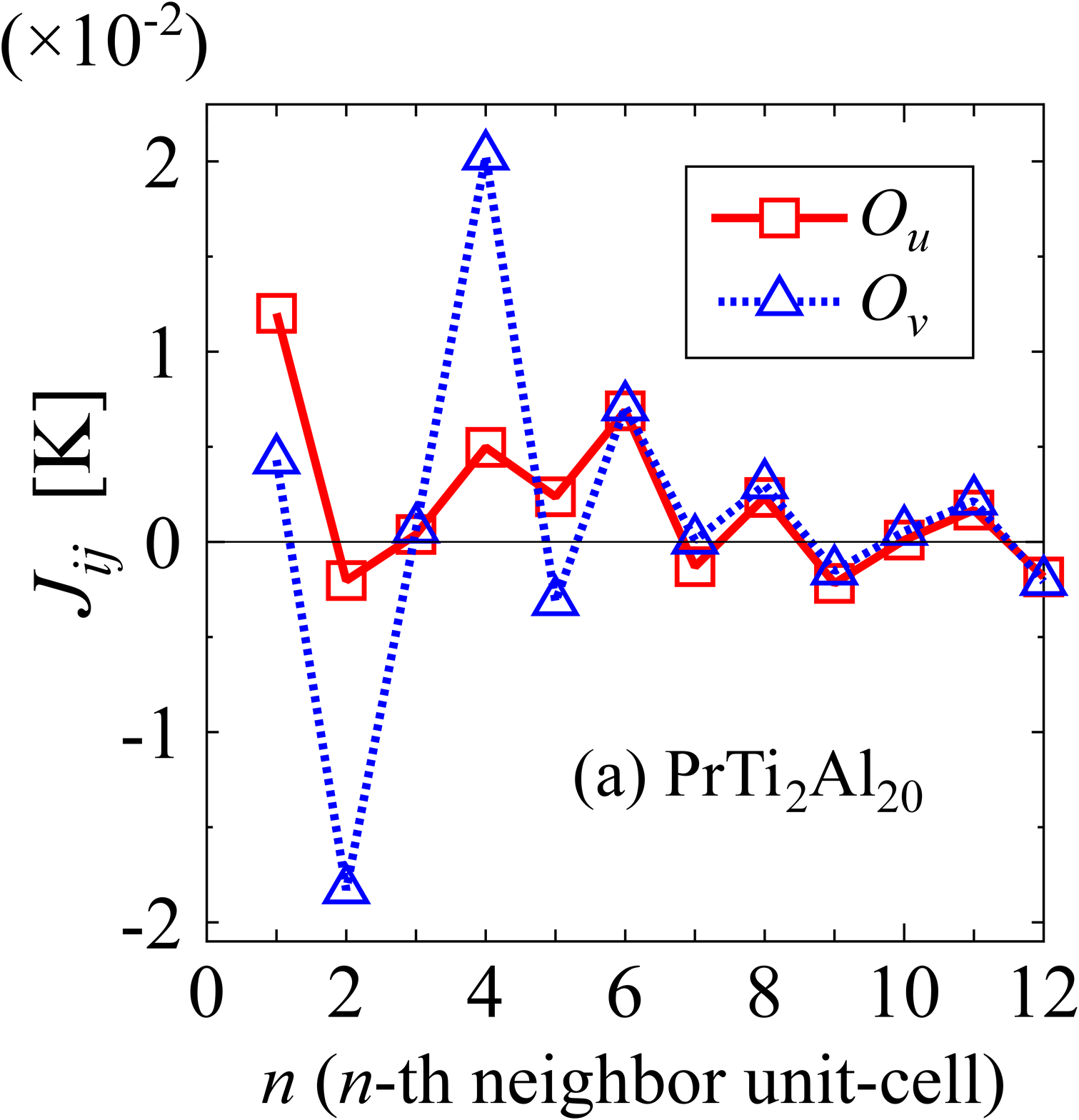}\vspace{5mm}
\includegraphics[width=90mm]{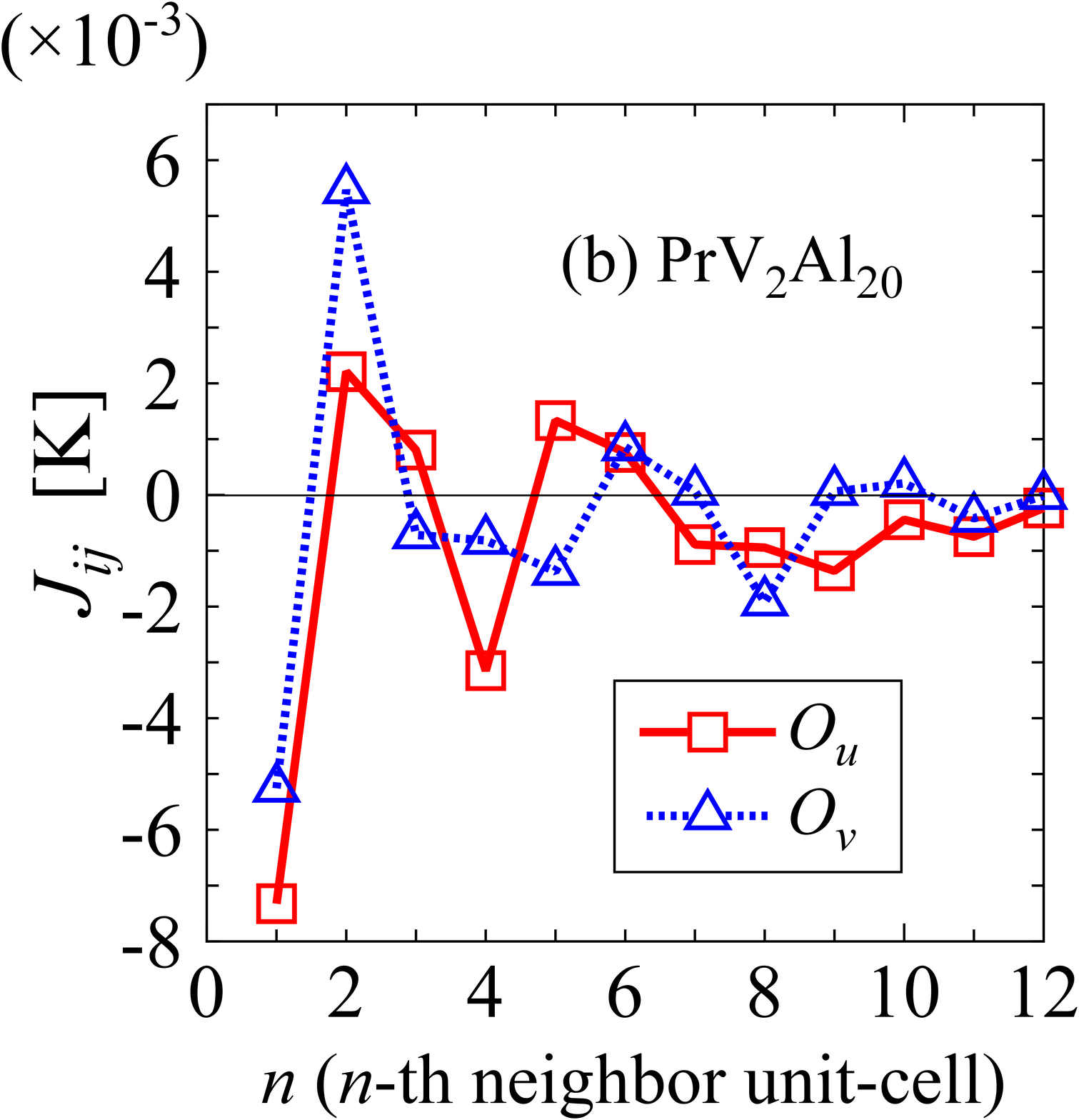}\vspace{2mm}
\caption{The RKKY interactions for LaTi$_{2}$Al$_{20}$ (a) and LaV$_{2}$Al$_{20}$ (b) as functions of the relative distance between $i$ and $j$.}
\label{f2}
\end{center}
\end{figure}

\begin{figure}[tbh]
\begin{center}
\includegraphics[width=90mm]{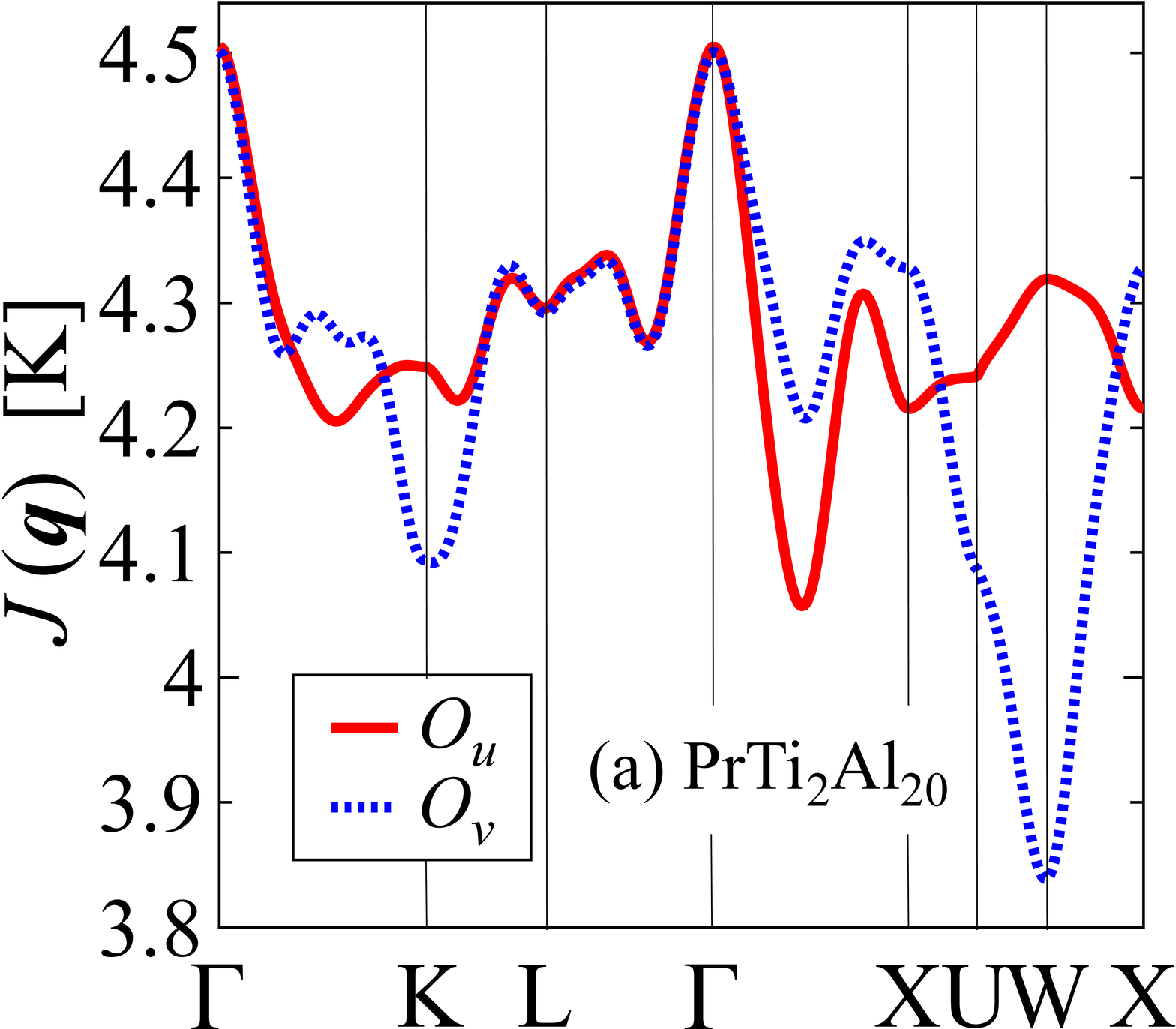}\vspace{7mm}
\includegraphics[width=90mm]{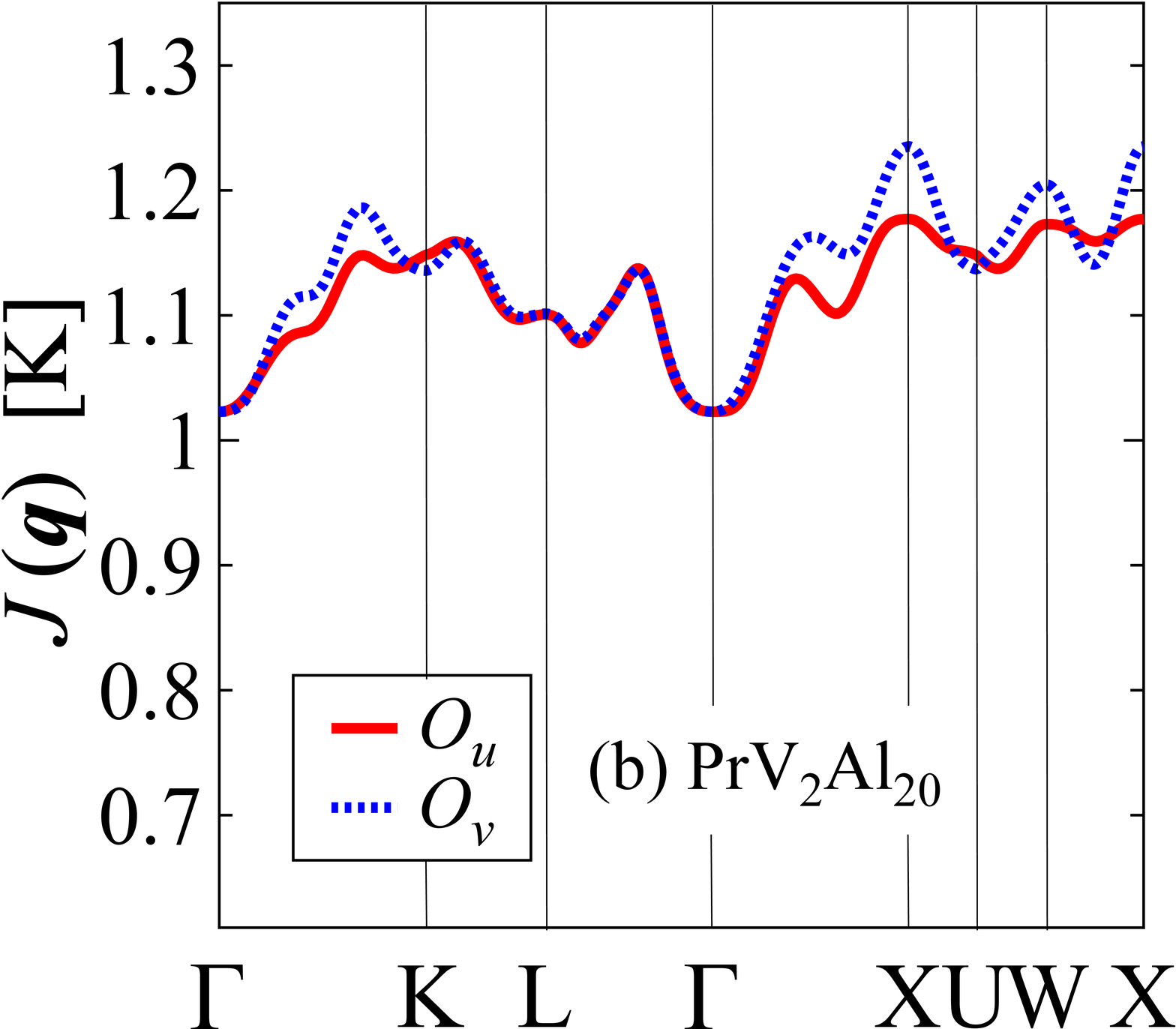}\vspace{3mm}
\caption{The Fourier transformed RKKY interactions for LaTi$_{2}$Al$_{20}$ (a) and LaV$_{2}$Al$_{20}$ (b) as functions of the weve vector, where $\Gamma=(0,0,0)$, ${\rm X}=(\pi/a,\pi/a,0)$, ${\rm U}=(5\pi/4a,5\pi/4a,\pi/2a)$, ${\rm W}=(\pi/a,3\pi/2a,\pi/2a)$, ${\rm K}=(3\pi/4a,3\pi/2a,3\pi/4a)$ and ${\rm L}=(\pi/a,\pi/a,\pi/a)$.}
\label{f3}
\end{center}
\end{figure}

\section*{Acknowledgments}
This work was partially supported by a Grant-in-Aid for Scientific Research from the Ministry of Education, Culture, Sports, Science and Technology. Numerical calculations were performed in part using OFP at the CCS, University of Tsukuba. 



\vspace{10mm}

[1] T. Onimaru and H. Kusunose, J. Phys. Soc. Jpn. {\bf 85}, 082002 (2016).

[2] M. Koseki {\it et al.}, J. Phys. Soc. Jpn. {\bf 80}, SA049 (2011).

[3] A. Sakai {\it et al.}, J. Phys. Soc. Jpn. {\bf 80}, 063701 (2011)

[4] A. Sakai {\it et al.}, J. Phys. Soc. Jpn. {\bf 81}, 083702 (2012)

[5] M. Matsushita {\it et al.}, J. Phys. Soc. Jpn. {\bf 80}, 074605 (2011)

[6] S. Nagashima {\it et al.}, JPS Conf. Proc. {\bf 3}, 011019 (2014)

[7] K. Hanzawa, J. Phys. Soc. {\bf 84}, 024717 (2015)

[8] M. J. Kangas {\it et al.}, J. Solid State Chem. {\bf 196}, 274281 (2012).

[9] K. Hanzawa, J. Phys. Soc. Jpn. {\bf 80}, 023707 (2011).


\end{document}